\documentclass[10pt,conference]{IEEEtran}
\usepackage{psfrag,graphicx,subfigure}
\usepackage{subfig}
\usepackage{cite,url}
\usepackage{stfloats}
\usepackage{amsfonts,amssymb,amsmath,amsbsy,bm,paralist,theorem,ifthen,color}
\usepackage{calc}
\usepackage{array,multirow,longtable}
\usepackage{algorithmic,algorithm}
\usepackage{graphics,booktabs,epsfig,enumerate}
\usepackage{tikz}
\usetikzlibrary{arrows}
\usepackage{tikz-3dplot}
\usepackage{cases}

\newtheorem{Proposition}{Proposition}

\newtheorem{Lemma}  {Lemma}

\theorembodyfont{\rmfamily}

\definecolor{orange}{RGB}{255,107,0}
\definecolor{green}{RGB}{0,180,80}

\newcommand{\Cset}  {\ensuremath{\mathbb{C}}}

\newcommand{\CN}    {\ensuremath{\mathcal{CN}}}

\newcommand{\hb}     {\ensuremath{\mathbf{h}}}
\newcommand{\vb}     {\ensuremath{\mathbf{v}}}
\newcommand{\xb}     {\ensuremath{\mathbf{x}}}

\newcommand{\zerob}  {\ensuremath{\mathbf{0}}}

\newcommand{\Hb}     {\ensuremath{\mathbf{H}}}

\newcommand{\Sb}     {\ensuremath{\mathbf{S}}}

\newcommand{\Xb}     {\ensuremath{\mathbf{X}}}

\DeclareMathOperator*{\st}{s.t.}
\DeclareMathOperator*{\rank}{rank}
\DeclareMathOperator*{\Tr}{Tr}

\begin{document}
\bibliographystyle{IEEEtran}
\title{Simultaneous Information and Energy Transfer: A Two-User MISO Interference Channel Case}
\author{Chao Shen$^\star$, Wei-Chiang Li$^\dag$ and Tsung-Hui Chang$^{\ddag}$\\[2pt]%
\begin{tabular}{c}
\small$^\star$Institute of Information Science, Beijing Jiaotong University, Beijing, China 100044\\
\small$^\dag$Institute of Commun. Eng., National Tsing Hua University, Hsinchu, Taiwan 30013\\
\small$^\ddag$Department of Elect. and Computer Eng., University of California, Davis, CA, USA 95616
\end{tabular}
\vspace{-0.1cm}}

\maketitle
\begin{abstract}
This paper considers the sum rate maximization problem of a two-user multiple-input single-output interference channel with receivers that can scavenge energy from the radio signals transmitted by the transmitters. We first study the optimal transmission strategy for an ideal scenario where the two receivers can simultaneously decode the information signal and harvest energy. Then, considering the limitations of the current circuit technology, we propose two practical schemes based on TDMA, where, at each time slot, the receiver either operates in the energy harvesting mode or in the information detection mode. Optimal transmission strategies for the two practical schemes are respectively investigated. Simulation results show that the three schemes exhibit interesting tradeoff between achievable sum rate and energy harvesting requirement, and do not dominate each other in terms of maximum achievable sum rate.
\end{abstract}
\begin{keywords}
  Energy harvesting, Energy transfer, Interference channel, Transmitter Optimization
\end{keywords}
\IEEEpeerreviewmaketitle

\vspace{-0.2cm}
\section{Introduction}\label{sec:intr}
\vspace{-0.1cm}

Recently, energy harvesting has been considered as a promising technique with great potential for prolonging the life time of the battery-powered mobile devices or for implementing self-sustained communication systems. Transmission designs with energy harvesting constraints imposed on the transmitter have been studied in \cite{Ozel2011,Zhang2011,Tutuncuoglu2012} (and references therein). Specifically, assuming that the transmitter is able to harvest energy from some external energy sources, the work in \cite{Ozel2011} investigated the optimal power allocation scheme for minimizing the transmission completion time in a point-to-point single-input and single-output (SISO) channel. In \cite{Zhang2011}, the throughput maximization problem was studied for a relay network with energy harvesting transmitters and relays. Transmission designs for an interference channel (IFC) with transmitter energy harvesting constraints were also studied in \cite{Tutuncuoglu2012}.

In some other works \cite{Varshney2008,Grover2010,Zhangrui2011}, on the other hand, the receivers were assumed to be able to scavenge energy from the radio signals transmitted by the transmitters. The assumption made there is that the receiver can simultaneously detect information bits and harvest energy from the received signal. Under this assumption, the works in \cite{Varshney2008} and \cite{Grover2010} investigated the optimal tradeoff between information and energy transfer in a SISO flat-fading channel and in a frequency-selective fading channel, respectively.
Considering the fact that simultaneous information detection and energy harvesting cannot be fulfilled by current circuit technologies, the work in \cite{Zhangrui2011} proposed two practical schemes where the receiver separates the modes for energy harvesting (EH) and information detection (ID) either over the time domain (i.e., TDMA) or over the power domain (i.e., power splitting). In \cite{Zhangrui2011}, the rate performance achieved by the ideal receiver which implements EH and ID simultaneously serves as an upper bound of the two practical schemes.

In this paper, we consider the transmission design problem for a two-user multiple-input single-output interference channel (MISO-IFC), assuming energy harvesting receivers. It is interesting to note that, despite that the cross-link signals are interference which limits the achievable sum rate, they are helpful in boosting the energy harvesting of the receivers. We first consider the ideal receivers which can simultaneously perform ID and EH. We formulate the design problem by maximizing the sum rate of the two transmitter-receiver pairs subject to \textit{minimum energy harvesting} constraints, i.e., constraints on the minimum amount of energy to be harvested. The considered problem is, however, intrinsically difficult to handle. We present an analysis to characterize the optimal solution structure, showing that transmit beamforming is an optimal strategy.

We further propose two practical schemes. The first scheme, which we call TDMA scheme A, divides the transmission time into two time slots. Both receivers perform EH in the first time slot and subsequently perform ID in the second time slot. The second scheme, which we call TDMA scheme B, again divides the transmission time into two time slots, but in each time slot, one receiver performs EH while the other performs ID. We respectively present efficient optimization methods for solving the transmission design problems associated with the two practical schemes.

Simulation results are presented to compare the achievable sum rates of the three proposed schemes. Intriguingly, we observe that the scheme with ideal receivers may not be ideal in terms of sum rate maximization. Instead, the practical TDMA scheme A may outperform the ideal scheme when the system is interference limited. Besides, the TDMA scheme B may also yield a higher sum rate than TDMA scheme A when one of the receivers requires much more energy than the other. We should mention that these results are very different from those in \cite{Zhangrui2011} where the ideal receiver always performs the best owing to the absence of interference.

\textit{Notations:} $\Tr(\Xb)$ represents the trace of matrix $\Xb$. $\Xb\!\succeq\!\zerob$ means that matrix $\Xb$ is positive semidefinite (PSD). $\|\xb\|$ denotes the Euclidean norm of vector $\xb$. The orthogonal projection onto the column space of a tall matrix $\Xb$ is denoted by $\Pi_\Xb\triangleq\Xb(\Xb^H\Xb)^{-1}\Xb^H$, and the projection onto the orthogonal complement of the column space of $\Xb$ is denoted by $\Pi_\Xb^\perp\triangleq\mathbf{I}-\Pi_\Xb$ where $\mathbf{I}$ is an identity matrix. 

\section{Signal Model and Problem Statement}\label{sec:sig_mod_prob_formu}

\tikzstyle{information text}=[text badly centered,font=\small,text
width=3cm]
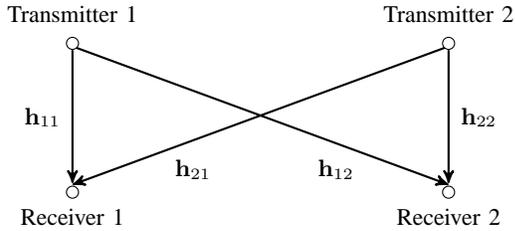
\begin{figure}[t]\centering
\beginpgfgraphicnamed{graphic-of-flat-world}
\begin{tikzpicture}[scale=1,cap=round]
    \begin{scope}[>=stealth']
        \draw[<-,thick] (0,0.12) -- (0,1.88);
        \draw[<-,thick] (5,0.12) -- (5,1.88);
        \draw[<-,thick] (0,0.1) -- (5,1.94);
        \draw[<-,thick] (5,0.1) -- (0,1.94);
    \end{scope}
    \draw (0, 2.6) node[below, information text] {Transmitter 1};
    \draw (5, 2.6) node[below, information text] {Transmitter 2};
    \draw (-0.4,1) node[information text]{$\hb_{11}$ };
    \draw (3.5,.3) node[information text]{$\hb_{12}$ };
    \draw (1.6,.3) node[information text]{$\hb_{21}$ };
    \draw (5.4,1)  node[information text]{$\hb_{22}$ };
    \draw (0,-0.1) node[below, information text] {Receiver 1};
    \draw (5,-0.1) node[below, information text] {Receiver 2};
\foreach \x in {0,5} \foreach \y in {0,2-0.02} {
\filldraw[fill=white] (\x,\y) circle (0.08); }
\end{tikzpicture}
\endpgfgraphicnamed
\caption{A two-user MISO-IFC system for simultaneous information and
energy transfer.} \label{Fig::systemmodel}\vspace{-0.5cm}
\end{figure}

We consider a two-user MISO-IFC, as shown in Fig. \ref{Fig::systemmodel}, where each transmitter is equipped with $N_t$ antennas and each receiver has a single antenna. Let $\xb_i \in \mathbb{C}^{N_t}$ denote the signal vector transmitted by transmitter $i$, and $\hb_{ik} \in \mathbb{C}^{N_t}$ denote the channel vector from transmitter $i$ to receiver $k$, for $i,k\in\{1,2\}$. The received signal at receiver $i$ is given by
\begin{align}\label{eq: received signal}
y_i = \hb_{ii}^H\xb_i+\hb_{ki}^H\xb_k+n_i,~k\neq i,
\end{align}
where $n_i\sim\CN(0,\sigma_i^2)$ is the additive Gaussian noise.

Different from the conventional MISO IFC \cite{Jorswieck08}, we assume in the paper that the receivers can either extract information or harvest energy from the received signal $y_i$ \cite{Varshney2008,Grover2010,Zhangrui2011}, which we refer to as the information detection (ID) mode and the energy harvesting (EH) mode, respectively.
Assume that $\xb_i$ contains the information intended for receiver $i$ which is Gaussian encoded with zero mean and covariance matrix $\Sb_i$, i.e., $\xb_i\sim\CN(\zerob,\Sb_i)$.
Moreover, assume that each receiver $i$ decodes $\xb_i$ by single user detection in the ID mode. Then the achievable information rates of the two receivers are respectively given by
\begin{align}
r_1(\Sb_1,\Sb_2)&=\log\left(1+\frac{\hb_{11}^H\Sb_1\hb_{11}}{\hb_{21}^H\Sb_2\hb_{21}
+ \sigma_1^2}\right),\\
r_2(\Sb_1,\Sb_2)&=\log\left(1+\frac{\hb_{22}^H\Sb_2\hb_{22}}{\hb_{12}^H\Sb_1\hb_{12}
+ \sigma_2^2}\right).
\end{align}

Alternatively, the receiver $i$ may choose to harvest energy from $y_i$, i.e., operating in the EH mode. In particular, it can be
assumed that the total harvested RF-band energy during a symbol transmission interval $\Delta$ is proportional to the power of the received baseband signal, e.g., for receiver $i$, the harvested energy, denoted by $\mathcal{E}_i$, can be expressed as
\begin{align}
  \mathcal{E}_i= \gamma \Delta (\hb_{1i}^H\Sb_1\hb_{1i} + \hb_{2i}^H\Sb_2\hb_{2i}),~i=1,2,
\end{align}
where $\gamma$ is a constant accounting for the energy conversion loss in the transducer \cite{Zhangrui2011}. It should be noted that current practical circuits do not allow the receiver to simultaneously decode the information bits and harvest the energy \cite{Zhangrui2011}.

Suppose that receiver $i$ desires to harvest a total amount of energy $E_i$ for $i=1,2$. Our interest lies in investigating the optimal transmission strategies of $\Sb_1$ and $\Sb_2$ so that the sum rate of the two transmitter-receiver pairs can be maximized while their energy harvesting requirements are satisfied at the same time. In the next section, we first study an `ideal' situation that the receiver can simultaneously operate in the ID mode and EH mode. In the subsequent sections, we further investigate two practical TDMA schemes where each receiver either operates in the ID mode or in the EH mode.

\section{Optimal Transmission Strategy for Ideal Receivers}\label{sec:opt_beamforming}

Let us first consider the ideal situation that the receiver can simultaneously decode the information bits and harvest the energy. Under such assumption, we consider the following problem formulation:
\begin{subequations}\label{(P)}
\begin{align}
{\rm (P)}~~\max_{\Sb_1\succeq\zerob,\,\Sb_2\succeq\zerob}~
    &r_1(\Sb_1,\Sb_2)+r_2(\Sb_1,\Sb_2)\label{(P)_a}\\
\st~&\hb_{11}^H\Sb_1\hb_{11} + \hb_{21}^H\Sb_2\hb_{21} \geq E_1,\label{(P)_b}\\
    &\hb_{22}^H\Sb_2\hb_{22} + \hb_{12}^H\Sb_1\hb_{12} \geq E_2,\label{(P)_c}\\
    &\Tr(\Sb_1) \leq P_1,\label{(P)_d}\\
    &\Tr(\Sb_2) \leq P_2,\label{(P)_e}
\end{align}
\end{subequations}
where \eqref{(P)_b} and \eqref{(P)_c} are the energy harvesting constraints (we have set $\gamma=\Delta=1$  for notational simplicity), and \eqref{(P)_d} and \eqref{(P)_e} are the individual power constraints.
Since, when $E_1=E_2=0$, problem (P) reduces to the well-known sum rate maximization problem in interference channels \cite{Jorswieck08}, (P) is difficult to solve in general . However, an explicit solution structure for $\Sb_1$ and $\Sb_2$ in (P) can be obtained, as we show in the following proposition:

\begin{Proposition}\label{proposition:(P)_rnk1}
Assume that (P) is feasible, and let $(\Sb_1^\star,\Sb_2^\star)$ denote the optimal solution pair of (P). Then $\Tr(\Sb_1^\star)=P_1$ and $\Tr(\Sb_2^\star)=P_2$. Moreover, there exists a pair of $(\Sb_1^\star,\Sb_2^\star)$ satisfying
\begin{subequations}\label{optimal S}
\begin{align}
   \Sb_1^\star= (a_1 \hb_{11} + b_1 \hb_{12})(a_1 \hb_{11} + b_1 \hb_{12})^H,\\
   \Sb_2^\star= (a_2 \hb_{21} + b_2 \hb_{22})(a_2 \hb_{21} + b_2 \hb_{22})^H,
\end{align}
\end{subequations}
for some $a_i,b_i \in \mathbb{C}$, $i=1,2$.
\end{Proposition}

Proposition \ref{proposition:(P)_rnk1} implies that beamforming is an optimal transmission strategy of (P), and that the beamforming direction of transmitter $i$ should lie in the range space of $[\hb_{i1}, \hb_{i2}]$, for $i=1,2$. By \eqref{optimal S}, the search of $\Sb_1$ and $\Sb_2$ in (P) reduces to the search of $a_i$ and $b_i$, over the ellipsoid $|a_i|^2 \|\hb_{i1}\|^2+|b_i|^2 \|\hb_{i2}\|^2=P_i$, for all $i=1,2.$ We should mention here that the optimal beamforming solution structure in \eqref{optimal S} is reminiscent of that in the traditional MISO IFC without energy harvesting constraints \cite{Jorswieck08},

\subsection{Proof of Proposition 1}

Without loss of generality, we assume that $\hb_{11}\nparallel\hb_{12}$ and $\hb_{21}\nparallel\hb_{22}$. We prove by contradiction that $\Tr(\Sb_i^\star)=P_i$ for $i=1,2$. Suppose that $\Tr(\Sb_1^\star)<P_1$, then there exists some
$\epsilon>0$ and
\[
\Sb_1'=\Sb_1^\star+\epsilon\Pi_{\hb_{12}}^\perp\hb_{11}\hb_{11}^H\Pi_{\hb_{12}}^\perp
\]
such that $\Tr(\Sb_1')=P_1$. Note that $(\Sb_1',\Sb_2^\star)$ is feasible to (P). Moreover, since $\hb_{11}\nparallel\hb_{12}$, we have $r_1(\Sb_1',\Sb_2^\star)>r_1(\Sb_1^\star,\Sb_2^\star)$ and $r_2(\Sb_1',\Sb_2^\star)=r_2(\Sb_1^\star,\Sb_2^\star)$, which contradicts the optimality of $(\Sb_1^\star,\Sb_2^\star)$. Hence, it must be that $\Tr(\Sb_1^\star)=P_1$; similarly, one can show that $\Tr(\Sb_2^\star)=P_2$.

Next, we show that $\Sb_1^\star$ and $\Sb_2^\star$ lie in the range space of $\Hb_1\triangleq[\hb_{11}~\hb_{12}]$ and $\Hb_2\triangleq[\hb_{21}~\hb_{22}]$, respectively, {i.e., $\Pi_{\Hb_i}^\perp \Sb_i^\star \Pi_{\Hb_i}^\perp=0$ for $i=1,2$.} One can see that, for
any $\Sb\succeq\zerob$,
\begin{align} \label{eq: equivalence}
 \hb_{ik}^H(\Pi_{\Hb_i}\Sb\,\Pi_{\Hb_i}^H)\hb_{ik}&=\hb_{ik}^H\Sb\hb_{ik},\\
 \Tr(\Pi_{\Hb_i}\Sb\,\Pi_{\Hb_i}^H)&\le\Tr(\Sb),
\end{align}
for $i,k\!\in\!\{1,2\}$, where the equality in \eqref{eq: equivalence} holds because $\Pi_{\Xb}\Xb\!=\!\Xb$ for all $\Xb\!\in\!\Cset^{m\times{n}}$. Therefore, $(\Sb_1^\star,\Sb_2^\star)$ is an optimal solution to problem (P) if and only if $(\Pi_{\Hb_1}\Sb_1^\star\Pi_{\Hb_1},\Pi_{\Hb_2}\Sb_2^\star\Pi_{\Hb_2})$ is optimal to problem (P). Now suppose that $\Sb_1^\star$ does not lie in the range space of $\Hb_1$, i.e., $\Tr(\Pi_{\Hb_1}^\perp\Sb_1^\star\Pi_{\Hb_1}^\perp)>0$. Then,
\[\Tr(\Pi_{\Hb_1}\Sb_1^\star\Pi_{\Hb_1}^H)=\Tr(\Sb_1^\star)-\Tr(\Pi_{\Hb_1}^\perp\Sb_1^\star\Pi_{\Hb_1}^\perp)<\Tr(\Sb_1^\star)\le{P_1},\]
which implies that $\Pi_{\Hb_1}\Sb_1^\star\Pi_{\Hb_1}$ is not optimal, and thereby $\Sb_1^\star$ is not optimal to (P). Analogously, one can show that $\Sb_2^\star$ must lie in the range space of $\Hb_2$.

What remains to prove \eqref{optimal S} is to show that there exists a pair of $(\Sb_1^\star,\Sb_2^\star)$ that are of rank one. It is not difficult to see that (P) is equivalent to the following problems
\begin{subequations}\label{opt_Si}
\begin{align}
\max_{\Sb_i\succeq\zerob}~
    &\log\left(1+\frac{\hb_{ii}^H\Sb_i\hb_{ii}}{\Gamma_{ki}^\star + \sigma_i^2}\right)\\
\st~&\hb_{ik}^H\Sb_i\hb_{ik} + \Gamma_{kk}^\star \geq E_k,\\
    &\Gamma_{ki}^\star + \hb_{ii}^H\Sb_i\hb_{ii} \geq E_i,\\
    &\hb_{ik}^H\Sb_i\hb_{ik}\leq \Gamma_{ik}^\star,\\
    &\Tr(\Sb_i) \leq P_i,
\end{align}
\end{subequations}
where $\Gamma_{ki}^\star\!=\!\hb_{ki}^H\Sb_k^\star \hb_{ki}$, $i,k\!\in\!\{1,2\}$ and $i\!\neq\!k$. Let us focus on the case of $i=1$, $k=2$, and rewrite \eqref{opt_Si} as
\begin{subequations}\label{opt_S1}
\begin{align}
\max_{\Sb_1\succeq\zerob}~
    &{\hb_{11}^H\Sb_1\hb_{11}}\label{opt_S1_a}\\
\st~&\hb_{12}^H\Sb_1\hb_{12} \geq E_2-\Gamma_{22}^\star,\label{opt_S1_b}\\
    &\hb_{12}^H\Sb_1\hb_{12} \leq \Gamma_{12}^\star,\label{opt_S1_d}\\
    &\hb_{11}^H\Sb_1\hb_{11} \geq E_1-\Gamma_{21}^\star,\label{opt_S1_c}\\
    &\Tr(\Sb_1) \leq P_1.\label{opt_S1_e}
\end{align}
\end{subequations}

Suppose that $\Gamma_{12}^\star=E_2-\Gamma_{22}^\star$. Then \eqref{opt_S1_b} and \eqref{opt_S1_d} merges to one equality constraint. In that case, \eqref{opt_S1} has only three inequality constraints. According to \cite[Theorem 3.2]{Huang2010}, problem \eqref{opt_S1} then has an optimal solution ${\Sb}_1^\star$ such that $\rank({\Sb}_1^\star)\le1$. On the other hand, if $\Gamma_{12}^\star>E_2-\Gamma_{22}^\star$, then one of the two constrains \eqref{opt_S1_b} and \eqref{opt_S1_d} must be inactive for $\Sb_1^\star$. Therefore, the effective number of inequalities in \eqref{opt_S1} is again three. Thus $\rank({\Sb}_1^\star)\le1$ by \cite[Theorem 3.2]{Huang2010}. Analogously, for the case of $i=2$, $k=1$, one can show that problem \eqref{opt_Si} has an optimal ${\Sb}_2^\star$ with $\rank({\Sb}_2^\star)\le 1$. The proof is thus complete. \hfill{$\blacksquare$}

It is important to remark that, while (P) is ideal in the sense that the receiver can simultaneously operate in the ID and EH modes, (P) may not be the optimal design formulation in terms of sum rate maximization. The reason is that the cross-link signal power $\hb_{ik}^H\Sb_i\hb_{ik}$, though boosting the energy harvesting of receiver $k$, also degrades the achievable information rate. Therefore, when the cross-link interference is strong (e.g., interference dominated scenario), it might be better to split the ID and EH modes, which is also preferred by the current practical circuits.

\section{Practical Schemes and Their Optimal Transmission Strategies}\label{sec:prac_trans}

In the section, based on the idea of TDMA, we present two practical schemes where each of the receivers operates either in the ID mode or in the EH mode. The associated optimal transmission strategies of $\Sb_1$ and $\Sb_2$ are also investigated.

\subsection{TDMA scheme A}\label{subsec:TDMA_A}

In the first practical scheme, which we call TDMA scheme A, the transmission interval is divided into two time slots-- one dedicated for the EH mode and the other for the ID mode.
Suppose that $\alpha$ fraction of the time is for time slot 1 and $(1-\alpha)$ fraction of the time is for time slot 2. TDMA scheme A is described as follows:
\begin{itemize}
\item Time slot 1 (EH mode): Both the two receivers operate in the EH mode. The goal is to guarantee the two receivers to achieve their respective energy harvesting requirements $E_1$ and $E_2$ in $\alpha$ fraction of the time, i.e.,
    \begin{subequations}\label{eq: energy requirement}
    \begin{align}
    &\alpha\cdot(\hb_{11}^H\Sb_1\hb_{11}+\hb_{21}^H\Sb_2\hb_{21})\ge{E_1},\\
        &\alpha\cdot(\hb_{22}^H\Sb_2\hb_{22}+\hb_{12}^H\Sb_1\hb_{12})\ge{E_2}.
    \end{align}
    \end{subequations}
\item Time slot 2 (ID mode): Both the two receivers operate in the ID mode. It is aimed to maximize the sum rate:
    \begin{subequations}\label{TDMA_A_ID}
    \begin{align}
    \max_{\Sb_1\succeq\zerob,\,\Sb_2\succeq\zerob}~&(1-\alpha)\left(r_1(\Sb_1,\Sb_2)+
    r_2(\Sb_1,\Sb_2)\right) \\
    \st~~&\!\!\!\Tr(\Sb_1)\leq P_1,~\Tr(\Sb_2)\leq P_2.
    \end{align}
    \end{subequations}
\end{itemize}

Problem \eqref{TDMA_A_ID} is the classical sum rate maximization problem in MISO IFC, and there exist several efficient algorithms for handling \eqref{TDMA_A_ID}; see, e.g., \cite{Lindblom2011,Jorswieck08}.

Since time slot 1 is only for energy harvesting and does not contribute to the information rate, it is desirable to spend as least as possible time for the EH mode, i.e., to minimize the time fraction $\alpha$. Mathematically, this can be expressed as the following design problem
\begin{subequations}\label{TDMA_A_EH}
\begin{align}
\max_{\beta\in \mathbb{R},\,\Sb_1\succeq\zerob,\,\Sb_2\succeq\zerob}&~~\beta\\
\st~~~&\!\!\hb_{11}^H\Sb_1\hb_{11}+\hb_{21}^H\Sb_2\hb_{21}\ge\beta{E_1},\\
&\!\!\hb_{12}^H\Sb_1\hb_{12}+\hb_{22}^H\Sb_2\hb_{22}\ge\beta{E_2},\\
&\!\!\!\Tr(\Sb_1)\le{P_1},~\Tr(\Sb_2)\le{P_2},
\end{align}
\end{subequations}
where $\beta\triangleq1/\alpha$. Note that if the optimal $\beta$ of \eqref{TDMA_A_EH} is less than one (i.e., optimal $\alpha> 1$), then it implies that the energy harvesting requirements \eqref{eq: energy requirement} cannot be fulfilled even the receivers operate in the EH mode for the whole symbol transmission period. In that case, we declare that TDMA scheme A is not feasible.

While problem \eqref{TDMA_A_EH} is
a linear program, which can be solved by the off-the-shelf solvers, we show next that the optimal solution of \eqref{TDMA_A_EH} can actually be obtained via solving a simple one-dimensional problem.

\begin{Proposition}\label{proposition: one dimensional prob}
Denote $(\beta^\star,\Sb_1^\star,\Sb_2^\star)$ as the optimal solution of \eqref{TDMA_A_EH}. Let
\begin{align}\label{golden_section}
w^\star=\arg~\max_{0\le w \le1}~\min\{\beta_1(w),\beta_2(w)\},
\end{align}
where
\begin{subequations}\label{TDMA_A_EH_beta}
\begin{align}
&\beta_1(w)=(P_1|\hb_{11}^H\vb_1(w)|^2+P_2|\hb_{21}^H\vb_2(w)|^2)/E_1,
   \label{TDMA_A_EH_beta_b}\\
&\beta_2(w)=(P_1|\hb_{12}^H\vb_1(w)|^2+P_2|\hb_{22}^H\vb_2(w)|^2)/E_2,
\label{TDMA_A_EH_beta_c}
\end{align}
\end{subequations}
and $\vb_1(w) \in \mathbb{C}^{N_t}$ and $\vb_2(w)\in \mathbb{C}^{N_t}$ are the principal eigenvectors of the two matrices
$w\hb_{11}\hb_{11}^H/E_1+(1-w)\hb_{12}\hb_{12}^H/E_2$ and
$w\hb_{21}\hb_{21}^H/E_1+(1-w)\hb_{22}\hb_{22}^H/E_2$, respectively. Then
\begin{subequations}\label{TDMA_A_EH_CLSFORM_SOL}
\begin{align}
&\Sb_1^\star=P_1\vb_1(w^\star)\vb_1^H(w^\star), \\
&\Sb_2^\star=P_2\vb_2(w^\star)\vb_2^H(w^\star), \label{TDMA_A_EH_CLSFORM_SOL_a}
\end{align}
\end{subequations} and $\beta^\star=\min\{\beta_1(w^\star),\beta_2(w^\star)\}$.
\end{Proposition}

\emph{\bf Proof:} Let us denote
\begin{subequations}\label{beta}
\begin{align}\label{beta1}
       &\beta_1=(\hb_{11}^H\Sb_1\hb_{11}+\hb_{21}^H\Sb_2\hb_{21})/E_1, \\
       &\beta_2=(\hb_{12}^H\Sb_1\hb_{12}+\hb_{22}^H\Sb_2\hb_{22})/E_2. \label{beta2}
\end{align}
\end{subequations} Then \eqref{TDMA_A_EH} can be written as
\begin{subequations}\label{eq: maxmin}
\begin{align}
\max_{\Sb_1\succeq\zerob,\,\Sb_2\succeq\zerob}&~~\min\{\beta_1,\beta_2\}\\
\st~~~&\!\!\text{constraints~in~}\eqref{beta1},~\eqref{beta2},\\
&\!\!\!\Tr(\Sb_1)\le{P_1},~\Tr(\Sb_2)\le{P_2}.
\end{align}
\end{subequations}
Define the following set
\begin{align}
\!\!\!\!\mathcal{P}\triangleq\Bigg\{(\beta_1,\beta_2)\bigg|\!\!\!\!
    \begin{array}{ll}
       &\eqref{beta1},~\eqref{beta2},\\
       &\Sb_i\succeq\zerob,\Tr(\Sb_i)\le{P_i},i=1,2,
    \end{array} \Bigg\}.
\end{align}
It can be verified that $\mathcal{P}$ is convex, and, moreover, the optimal tuple $(\beta_1,\beta_2)$ of \eqref{eq: maxmin} must lie on the Pareto boundary of $\mathcal{P}$.
Therefore, there must exist a value $0\leq w^\star\leq 1$ such that $(\Sb_1^\star,\Sb_2^\star)$ of \eqref{eq: maxmin} is also optimal to the following problem
\begin{subequations}\label{TDMA_A_EH_weighted}
\begin{align}
\max_{\substack{\Sb_1\succeq\zerob,\Sb_2\succeq\zerob}}~&w^\star\beta_1+(1-w^\star)\beta_2\\
\st~&\text{constraints~in~}\eqref{beta1},~\eqref{beta2},\\
&\!\!\!\Tr(\Sb_1)\le{P_1},~\Tr(\Sb_2)\le{P_2}.
\end{align}
\end{subequations}
By \eqref{TDMA_A_EH_weighted} and by \cite[Proposition 2.1]{Zhangrui2011},  $(\Sb_1^\star,\Sb_2^\star)$ is thus given by the forms in \eqref{TDMA_A_EH_CLSFORM_SOL}.
By substituting \eqref{TDMA_A_EH_CLSFORM_SOL} into \eqref{beta}, defining $\beta_1(w^\star)$ and $\beta_2(w^\star)$ as in \eqref{TDMA_A_EH_beta}, and by \eqref{eq: maxmin}, we obtain the optimal objective value of \eqref{TDMA_A_EH} as
$$\beta^\star=\min\{\beta_1(w^\star),\beta_2(w^\star)\}.$$

Since for any $0\leq w\leq 1$, the corresponding solution of \eqref{TDMA_A_EH_weighted} is feasible to \eqref{eq: maxmin}, the optimal $w^\star$ is given by \eqref{golden_section}. \hfill $\blacksquare$

One can further show that the function $\min\{\beta_1(w),\beta_2(w)\}$ in \eqref{golden_section} is \textit{unimodal}, and thus \eqref{golden_section} can be efficiently solved by the bisection or golden search methods. Due to space limitations, we
omit the proof here; it will be presented in our future publication.
Note that Proposition 2 also implies that beamforming is optimal to the TDMA scheme A.

\subsection{TDMA scheme B}\label{subsec:TDMA_B}
Different from TDMA scheme A, in each time slot of TDMA scheme B, one receiver operates in the ID mode and the other receiver operates in the EH mode. Specifically,
\begin{itemize}
\item Time slot 1: Receiver 1 operates in the ID mode and receiver 2 operates in the EH mode. The objective is to maximize the information rate of receiver 1 while guaranteeing the energy harvesting requirement of receiver 2. The design problem is given by
\begin{subequations}\label{TDMA_B_1}
\begin{align}
\max_{\Sb_1\succeq\zerob,\,\Sb_2\succeq\zerob}~&\alpha\log\left(1+\frac{\hb_{11}^H\Sb_1\hb_{11}}{\hb_{21}^H\Sb_2\hb_{21}+\sigma_1^2}\right)\label{TDMA_B_1_a}\\
\st~&\hb_{12}^H\Sb_1\hb_{12}+\hb_{22}^H\Sb_2\hb_{22}\ge{E_2}/\alpha,\label{TDMA_B_1_b}\\
&\Tr(\Sb_1)\le{P_1},~\Tr(\Sb_2)\le{P_2},\label{TDMA_B_1_c}
\end{align}
\end{subequations}
\item Time slot 2: The operation modes of the two receivers are exchanged:
\begin{subequations}\label{TDMA_B_2}
\begin{align}
\max_{\Sb_1\succeq\zerob,\,\Sb_2\succeq\zerob}~&(1-\alpha)\log\left(1+\frac{\hb_{22}^H\Sb_2\hb_{22}}{\hb_{12}^H\Sb_1\hb_{12}+\sigma_2^2}\right)\\
\st~~~~&\!\!\!\!\hb_{11}^H\Sb_1\hb_{11}+\hb_{21}^H\Sb_2\hb_{21}\!\ge\!{E_1}/(1-\alpha),\\
&\!\!\!\Tr(\Sb_1)\le{P_1},~\Tr(\Sb_2)\le{P_2}.
\end{align}
\end{subequations}
\end{itemize}

First of all, it is easy to see that

\begin{Lemma}
 The TDMA scheme B is feasible if and only if
\begin{align}\label{TDMA-B_E-limit}
&\frac{E_1}{P_1\|\hb_{11}\|^2+P_2\|\hb_{21}\|^2}+\frac{E_2}{P_1\|\hb_{12}\|^2+P_2\|\hb_{22}\|^2}
\le 1.
\end{align}
\end{Lemma}

\emph{\bf Proof:} TDMA scheme B is feasible if and only if both \eqref{TDMA_B_1} and \eqref{TDMA_B_2} are feasible. Problem \eqref{TDMA_B_1} is feasible if and only if there exists some $\alpha\in[0,1]$ such that
\begin{align}
E_2&\le\alpha\cdot
\begin{pmatrix}
  \max\limits_{\Sb_1\succeq\zerob,\Sb_2\succeq\zerob}&\!\!\!\hb_{12}^H\Sb_1\hb_{12}+\hb_{22}^H\Sb_2\hb_{22}\\
  {^{\Tr(\Sb_1)\le{P_1},\Tr(\Sb_2)\le{P_2}}}&
\end{pmatrix} \notag \\
&=\alpha\cdot(P_1\|\hb_{12}\|^2+P_2\|\hb_{22}\|^2), \label{temp1}
\end{align} where the equality is obtained by \cite[Proposition 2.1]{Zhangrui2011}.
Similarly, one can show that \eqref{TDMA_B_2} is feasible if and only if
\begin{align}
E_1\le
(1-\alpha)\cdot(P_1\|\hb_{11}\|^2+P_2\|\hb_{21}\|^2).\label{temp2}
\end{align}
Combining \eqref{temp1} and \eqref{temp2} gives rise to \eqref{TDMA-B_E-limit}. Conversely, given \eqref{TDMA-B_E-limit}, one can show that there exists $\alpha\in[0,1]$ such that \eqref{temp1} and \eqref{temp2} hold with equalities. \hfill $\blacksquare$

Problem \eqref{TDMA_B_1} and problem \eqref{TDMA_B_2} are quasi-convex problems. While quasi-convex problems can be solved by the bisection technique, we show that \eqref{TDMA_B_1} and problem \eqref{TDMA_B_2} can actually be recast as convex problems, by applying the Charnes-Cooper transformation \cite{Charnes1962}. To illustrate this, let us take \eqref{TDMA_B_1} as an example.
Consider the following convex semidefinite program (SDP)
\begin{subequations}\label{CC_trans_2}
\begin{align}
\max_{\Xb_1\succeq\zerob,\,\Xb_2\succeq\zerob,\,y\ge0}~&\alpha\log\left(1+\hb_{11}^H\Xb_1\hb_{11}\right)\\
\st~&\hb_{21}^H\Xb_2\hb_{21}+y\sigma_1^2=1, \label{CC_trans_2 b}\\
&\hb_{12}^H\Xb_1\hb_{12}+\hb_{22}^H\Xb_2\hb_{22}\ge{yE_2/\alpha},\\
&\Tr(\Xb_1)\le{yP_1},~\Tr(\Xb_2)\le{yP_2},
\end{align}
\end{subequations}
Note that the optimal $y^\star$ of \eqref{CC_trans_2} must be positive; otherwise we have $\Xb^\star_1=\Xb_2^\star=\zerob$ which violates \eqref{CC_trans_2 b}.
Moreover, consider the following correspondence:
\begin{subequations}\label{transform}
\begin{align}
 &y=1/(\hb_{21}^H\Sb_2\hb_{21}+\sigma_1^2)>0, \\
 &\Xb_1=y\Sb_1,~\Xb_2=y\Sb_2.
\end{align}
\end{subequations}
Then, one can show that $(\Sb_1,\Sb_2)$ is feasible to \eqref{TDMA_B_1} if and  only if $(\Xb_1,\Xb_2,y)$ is feasible to \eqref{CC_trans_2}. Furthermore, the objective value achieved by $(\Sb_1,\Sb_2)$ in \eqref{TDMA_B_1} is the same as the objective value achieved by $(\Xb_1,\Xb_2,y)$ in \eqref{CC_trans_2}.
Therefore, the two problems \eqref{TDMA_B_1} and \eqref{CC_trans_2} are equivalent, and one actually can obtain $(\Sb_1^\star,\Sb_2^\star)$ of \eqref{TDMA_B_1} by solving the convex problem \eqref{CC_trans_2}. In addition, by applying \cite[Theorem 3.2]{Huang2010}, one can show that \eqref{CC_trans_2} has rank-one optimal $(\Xb_1^\star,\Xb_2^\star)$, implying that beamforming is also optimal to the TDMA scheme B.

It is also possible to obtain a closed-form solution to problem \eqref{TDMA_B_1}, if the channel condition favors receiver 2 to harvest the energy:

\begin{Lemma}\label{proposition:TDMA_B_closed_form}
Consider problem \eqref{TDMA_B_1} and assume that
\begin{align}\label{channel condition}
P_1|\hb_{12}\hat{\hb}_{11}|^2+P_2|\hb_{22}^H\hat{\hb}_{21\perp}|^2\ge{E_2/\alpha},
\end{align}
where $\hat{\hb}_{11}\triangleq{\hb_{11}}/{\|\hb_{11}\|}$ and $\hat{\hb}_{21\perp}\triangleq{\Pi_{\hb_{21}}^\perp\hb_{22}}
/{\|\Pi_{\hb_{21}}^\perp\hb_{22}\|}$.
Then $(\Sb_1^\star,\Sb_2^\star)=(P_1\hat{\hb}_{11}\hat{\hb}_{11}^H,P_2\hat{\hb}_{21\perp}\hat{\hb}_{21\perp}^H)$ is optimal to \eqref{TDMA_B_1}.
\end{Lemma}

\emph{\bf Proof:} Consider the following optimization problem
\begin{subequations}\label{TDMA_B_NoETConstr}
\begin{align}
\max_{\Sb_1\succeq\zerob,\,\Sb_2\succeq\zerob}~&\log\left(1+\frac{\hb_{11}^H\Sb_1\hb_{11}}{\hb_{21}^H\Sb_2\hb_{21}+\sigma_1^2}\right)\\
\st~&\Tr(\Sb_1)\le{P_1},~\Tr(\Sb_2)\le{P_2},
\end{align}
\end{subequations}
which is obtained by removing \eqref{TDMA_B_1_b} from \eqref{TDMA_B_1}. Since the objective function is strictly increasing w. r. t. $\hb_{11}^H\Sb_1\hb_{11}$ and strictly decreasing w. r. t. $\hb_{21}^H\Sb_2\hb_{21}$, it can be easily seen that $(\Sb_1^\star,\Sb_2^\star)=(P_1\hat{\hb}_{11}\hat{\hb}_{11}^H,P_2\hat{\hb}_{21\perp}\hat{\hb}_{21\perp}^H)$ is optimal to \eqref{TDMA_B_NoETConstr}. Substituting this $(\Sb_1^\star,\Sb_2^\star)$ into \eqref{TDMA_B_1_b}, we see that $(\Sb_1^\star,\Sb_2^\star)$ is also feasible to \eqref{TDMA_B_1} owing to the premise of \eqref{channel condition}. Consequently, $(\Sb_1^\star,\Sb_2^\star)$ is also optimal to \eqref{TDMA_B_1}.\hfill{$\blacksquare$}

\section{Simulation Results}

In this section, we present some simulation results to compare the three transmission schemes, namely, problem (P), TDMA scheme A and TDMA scheme B. We assume that each transmitter has two antennas ($N_t\!=\!2$), and set $P_1\!=\!P_2\!=\!1$ and $\sigma^2\triangleq \sigma_1^2=\sigma_2^2$. The channel vectors are randomly generated following complex Gaussian distribution. Specifically, we will present simulation comparison results of the following two sets of channel realizations: \\
{\bf Channel realization 1:}
\begin{align*}
  &\hb_{11}=
  \begin{bmatrix}
   0.0608 - 0.1896j \\
    -0.4942 - 0.1212j
  \end{bmatrix},
  \hb_{12}=
  \begin{bmatrix}
    0.7306 - 0.6496j \\
     -0.0369 - 0.1672j
  \end{bmatrix}, \notag \\
  &\hb_{21}=
  \begin{bmatrix}
    -0.4320 - 0.3112j \\
     -0.4142 - 0.0515j
  \end{bmatrix},
  \hb_{22}=
  \begin{bmatrix}
    0.5634 + 0.2935j \\
     -0.0672 - 0.2515j
  \end{bmatrix},
\end{align*}
where $j\!=\!\sqrt{-1}$. The norms of the channel vectors are $\|\hb_{11}\|=0.5464$,  $\|\hb_{12}\|=0.9925$, $\|\hb_{21}\|=0.6765$, $\|\hb_{22}\|=0.6865$, respectively. The noise variance $\sigma^2$ is set to $0.1$.\\
{\bf Channel realization 2:} The direct-link channels $\hb_{11}$ and $\hb_{22}$ are the same as those for Channel realization 1, and the cross-link channels are given by
\begin{align*}
  \hb_{12}=
  \begin{bmatrix}
    0.8948 - 0.7956j \\
     -0.0452 - 0.2047j
  \end{bmatrix}\!,~
  &\hb_{21}=
  \begin{bmatrix}
    -0.5291 - 0.3811j \\
     -0.5073 - 0.0630j
  \end{bmatrix},
\end{align*} whose norms are $\|\hb_{12}\|=1.2156$ and $\|\hb_{21}\|=0.8286$. The noise variance $\sigma^2$ is set to $0.001$. One can see that, for Channel realization 2, interference will be the major factor that limits the sum rate.

In the simulations, both problem (P) and problem \eqref{TDMA_A_ID} for TDMA scheme A are solved by an exhaustive search method similar to that in \cite{Lindblom2011}. The optimal time fraction $\alpha$ of TDMA scheme A is solved by Proposition 2. For TDMA scheme B, the associated optimal time fraction $\alpha$ is obtained via exhaustive search over $[0,1]$. The SDP problem \eqref{CC_trans_2} for both time slot 1 and slot 2 are solved by \texttt{CVX} \cite{CVX}.

\begin{figure*}[t]\centering%
  {\subfigure[Problem (P) versus TDMA scheme A]
  {\includegraphics[width=0.45\linewidth]{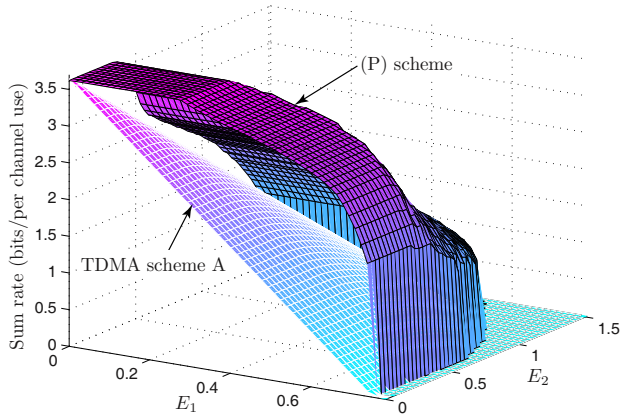}}}%
  \hspace{0.3cm}
  {\subfigure[TDMA scheme A versus TDMA scheme B]
  {\includegraphics[width=0.45\linewidth]{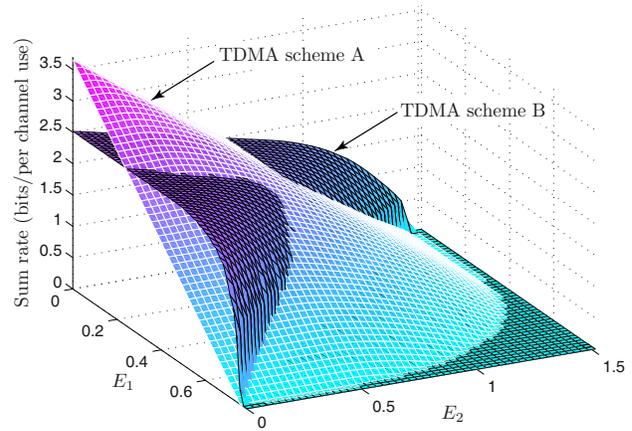}}}%
  \caption{Sum rate versus $(E_1,E_2)$ under Channel realization 1.}
  \label{fig1:EEregion}
\end{figure*}

In Figure \ref{fig1:EEregion}, we present the simulation results of sum rate versus $(E_1,E_2)$ of the three transmission schemes under Channel realization 1. Figure \ref{fig1:EEregion}(a) displays the comparison results between problem (P) and TDMA scheme A; while Figure \ref{fig1:EEregion}(b) shows the comparison results between TDMA scheme A and TDMA scheme B. We can observe from Figure \ref{fig1:EEregion}(a) that problem (P), which ideally assumes that the receivers can simultaneously decode the information bits and harvests the energy, exhibits a higher sum rate then TDMA scheme A for all values of $(E_1,E_2)$. Note that, when $(E_1,E_2)=(0,0)$, the two schemes coincides, thus they have the same sum rate at that point. From Figure \ref{fig1:EEregion}(b), we see that the two practical schemes, TDMA scheme A and TDMA scheme B, do not dominate each other in terms of sum rate, though TDMA scheme A has a higher sum rate for most of the values of $(E_1,E_2)$. It is interesting to see that TDMA scheme B performs better when either of the two receivers requests a higher energy than the other.

\begin{figure*}[t]\centering%
  {\subfigure[Problem (P) versus TDMA scheme A]
  {\includegraphics[width=0.45\linewidth]{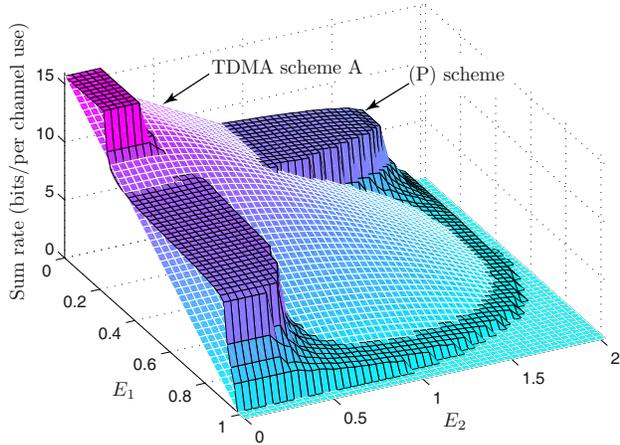}}}%
  \hspace{0.3cm}
  {\subfigure[TDMA scheme A versus TDMA scheme B]
  {\includegraphics[width=0.45\linewidth]{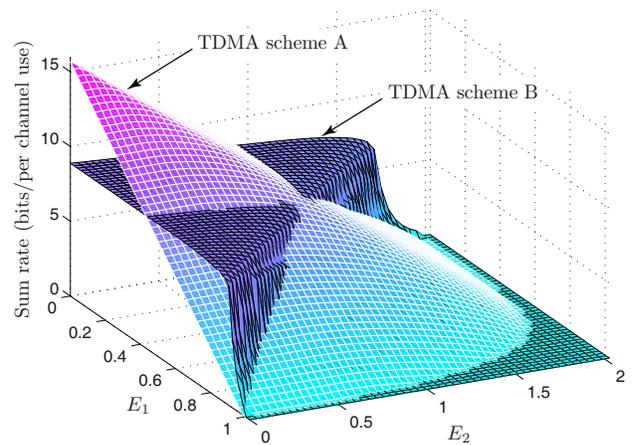}}}%
  \caption{Sum rate versus $(E_1,E_2)$ under Channel realization 2.}
  \label{fig1:EEregion2}\vspace{-0.4cm}
\end{figure*}

Figure \ref{fig1:EEregion2} presents the simulation results under Channel realization 2. Comparing Figure \ref{fig1:EEregion2}(a) with Figure \ref{fig1:EEregion}(a), one can observe that, in the interference-dominated scenario, TDMA scheme A may even yield a higher sum rate than problem (P). This implies that the `ideal' formulation (P) may not be `ideal' in terms of sum rate maximization when interference dominates the system performance. Similarly, we see from Figure \ref{fig1:EEregion2}(b) that TDMA scheme B can outperform TDMA scheme A when $E_1 >> E_2$ or $E_2 >> E_1$; otherwise TDMA scheme A exhibits a higher sum rate.

In summary, we have investigated the transmission design problem for simultaneous information and energy transfer in a two-user MISO interference channel. We have proposed three operation schemes, namely, the ideal problem (P), and two practical schemes -- TDMA scheme A and TDMA scheme B. We have analyzed the solution structures of the three schemes, showing that beamforming is optimal for the three schemes. Efficient methods for handling the design problems for the TDMA scheme A and TDMA scheme B are also presented. Simulation results have shown that the three schemes do not dominate each other in terms of sum rate. Future works will analytically compare the three schemes, and extend the framework to a general $K$-user interference channel.

\vspace{-0.0cm} \footnotesize

\end{document}